\documentclass[final,3p,times,sort&compress]{elsarticle}
\usepackage{
 array
,booktabs
,graphicx
,hyperref
,subfig
}
\usepackage[thinspace,thinqspace,squaren,textstyle]{SIunits}
\usepackage{
 amsmath
,amssymb
,braket
,feynmp
,slashed
}
\usepackage{
 amsfonts
,dsfont
,mathrsfs
}
\newcommand{\abs}[1]{\lvert#1\rvert}
\newcommand{\ep}[0]{\epsilon}
\newcommand{\s}[1]{\slashed{#1}}
\newcommand{\T}[1]{\operatorname{T}\mathopen{}\left\lbrace#1\mathclose{}\right\rbrace}
\DeclareMathOperator{\B}{\mathscr{B}_{\mathnormal{M}^2}}
\journal{arXiv.org}
\begin{document}%
\begin{frontmatter}%
\title{The nucleon wave function at the origin}
\author{Michael Gruber}
\address{
  Institut f{\"u}r Theoretische Physik, Universit{\"a}t Regensburg\\
  93040 Regensburg, Germany
}
\date{\today}%
\begin{abstract}%
  We calculate the next-to-leading order perturbative corrections to the SVZ sum rules for the coupling $f_N$, the nucleon leading twist wave function at the origin. The results are compared to the established Ioffe sum rules and also to lattice QCD simulations.
\end{abstract}%
\end{frontmatter}%
\section{\label{sec:intro}Introduction}%
Hard exclusive reactions have long been recognized as
an important tool in the exploration of the nucleon structure at
different scales. The emergence of quarks and gluons as the
adequate degrees of freedom is expected to happen at momentum
transfers that are accessible in present and planned experiments,
in particular on nucleon electromagnetic form factors and electroproduction
of nucleon resonances see e.g.\
\cite{Punjabi:2005wq,Gayou:2001qd,Burkert:2008mw,Aznauryan:2009da}.

Quantum chromodynamics (QCD) predicts 
\cite{Chernyak:1977as,Efremov:1979qk,Lepage:1979za,Lepage:1980fj}
that at large momentum transfer the form factors become increasingly
dominated by the contribution of the valence Fock state with small transverse
separation between the partons. In the collinear approximation the
wave function can be written in terms of the momentum fraction
distribution of the three valence quarks $ \varPhi_3(x_1,x_2,x_3)$,
dubbed distribution amplitude 
\begin{align}%
\begin{split}%
\ket{P_\uparrow}_{1/2} &= \frac{1}{24} \int \frac{[dx]}{\sqrt{x_1x_2x_3}} \varPhi_3(x_1,x_2,x_3)
\\
\times\ep^{abc} &u^\dagger_{a\downarrow}(x_1) \Bigl\lbrace u^\dagger_{b\downarrow}(x_2) d^\dagger_{c\uparrow}(x_3) - d^\dagger_{b\downarrow}(x_2) u^\dagger_{c\uparrow}(x_3) \Bigr\rbrace \ket{0} ~,
\end{split}%
\end{align}%
where the integration measure for the momentum fractions is defined as
$[dx]=dx_1dx_2dx_3\delta(x_1+x_2+x_3-1)$ 
and arrows indicate quark helicities. The integral of the distribution amplitude
defines a dimensionful constant
\begin{align}%
  f_N = \int [dx] \, \varPhi_3(x_1,x_2,x_3) ~,
\end{align}%
which determines the nucleon wave function at the origin. It is a fundamental 
(scale-dependent) nonperturbative constant which plays the central role 
in QCD description of hard exclusive reactions with protons, 
and determines an overall
normalization of the amplitudes. In the academic limit of very large 
momentum transfers the shape of the distribution amplitude 
$ \varPhi_3(x_1,x_2,x_3)$ is fixed by asymptotic freedom 
\cite{Lepage:1979za,Lepage:1980fj} and e.g.\ the 
proton magnetic form factor is determined entirely by $f_N$.

This constant has been estimated several times in the past using QCD sum rules
\cite{Chernyak:1984bm,King:1986wi,Chernyak:1987nv} and more recently
also in lattice QCD simulations \cite{Braun:2008ur}.  Different QCD sum rule
calculations are consistent with each other but the lattice result
appears to be $30\%$ lower. This discrepancy calls for a reevaluation of QCD
sum rules for $f_N$ including higher-order (NLO) contributions,
which is the goal of this work. Another motivation to derive the QCD sum rule
for $f_N$ to NLO accuracy is that it enters calculations of baryon 
form factors using the light-cone sum rule approach
\cite{Braun:2001tj,Braun:2006hz}
which are currently being advanced to the NLO as well
\cite{PassekKumericki:2008sj}.

The presentation is organized as follows:
In Sec.~\ref{sec:currents} we introduce two relevant interpolating currents for the nucleon and define the corresponding normalization constants. The general properties of their two-point functions as well as the diagrams contributing to them are presented in Sec.~\ref{sec:twopoint}. The results of our calculations are listed in Sec.~\ref{sec:sumrules}, followed by a numerical analysis in order to extract a new value for $f_N$ in Sec.~\ref{sec:analysis}. We conclude in Sec.~\ref{sec:conclusion}.
\section{\label{sec:currents}Proton interpolating currents}%
The leading twist normalization constant $f_N$ can be defined via a nucleon matrix element of a local three-quark operator:
\begin{align}%
\begin{split}%
\label{eq:CZ}
  \eta_{CZ}(x) = \frac{2}{3} \ep^{abc} \Bigl[ &\Bigl({u^a}^T(x)C\s{z} u^b(x)\Bigr) \gamma_5\s{z} d^c(x)
\\
  -&\Bigl({u^a}^T(x)C\s{z} d^b(x)\Bigr) \gamma_5\s{z} u^c(x) \Bigr] ~.
\end{split}%
\end{align}%
Here, $a,b,c$ are color indices, $C$ is the charge conjugation matrix ($C\gamma_\mu^TC^{-1}=-\gamma_\mu$) and $z$ is an arbitrary light-cone vector ($z^2=0$). The prefactor $2/3$ has been chosen to obtain the normalization given below in \eqref{eq:normalization}. Note that $\eta_{CZ}$ is constructed such that only isospin-$1/2$ contributions emerge when acting on a baryon state.
In what follows we refer to \eqref{eq:CZ} as the
(isospin improved) Chernyak-Zhitnitsky (CZ) current\ \cite{Chernyak:1984bm}.

For comparison we will also consider sum rules for another operator, known as
Ioffe current\ \cite{Ioffe:1981kw}:
\begin{align}%
  \eta_I(x) = \ep^{abc} \Bigl({u^a}^T(x)C\gamma_\mu u^b(x)\Bigr) \gamma_5\gamma^\mu d^c(x) ~.
\end{align}%
The matrix elements of these operators between vacuum and the proton state define the 
coupling constants, in standard notation:
\begin{subequations}%
\begin{align}%
\label{eq:normalization}
  \bra{0} \eta_{CZ}(0) \ket{P(q)} &= f_N (qz) \s{z} N ~,
\\
  \bra{0} \eta_{I}(0)  \ket{P(q)} &= \lambda_1 m_N N ~.
\end{align}%
\end{subequations}%
Here $N$ is a nucleon (proton) spinor 
and $m_N\simeq\unit{938}{\mega\electronvolt}$ the nucleon mass,  $q^2=m_N^2$.
In our calculations the three quark fields in the currents will be taken to be massless. For an analysis
of the properties of baryonic currents containing two massless and one massive quark see \cite{Groote:2008dx}.

To avoid confusion we note that the choice of the current is determined by the purpose 
of the calculation. In most QCD sum rule calculations of nucleon properties the Ioffe current
is usually adopted as a standard choice since it produces stable results. In these applications 
the value of the coupling $\lambda_1$ is not interesting by itself and usually 
cancels out in final results (e.g.\ in sum rule ratios). 
In our case, however, it is the coupling $f_N$ itself which is of interest,
so that we are bound to use the CZ current.
\section{\label{sec:twopoint}Two-point functions}%
The starting point for our calculation is a two-point function $\varPi(q)$ for a 
generic baryonic current $\eta(x)$:
\begin{align}%
\label{eq:2point}
  \varPi(q) = i\int d^4x \, e^{iqx} \bra{0}\T{ \eta(x) \bar{\eta}(0) }\ket{0} ~,
\end{align}%
with $\operatorname{T}$ denoting a time-ordered product and $\ket{0}$ the vacuum.

The two-point function can be decomposed into two parts via
\begin{align}%
  \varPi(q) = \s{q}\varPi_1(q^2) + \mathds{1}\varPi_2(q^2) ~.
\end{align}%
From dimension counting it becomes clear that $\varPi_1$ can only contain contributions from even-dimensional operators, whereas all odd-dimensional ones are contained in $\varPi_2$. 

We will consider the following contributions to $\varPi_1$: The unity operator, the two-gluon condensate and four-quark condensates. They will be calculated in perturbation theory up to next-to-leading order. For the Ioffe current the next-to-leading order contributions have already been calculated, e.g.\ in \cite{Jamin:1987gq,Ovchinnikov:1991mu} and the corresponding sum rules have been analyzed in \cite{Sadovnikova:2005ye}. For the Chernyak-Zhitnitsky current however, only the condensates are known at $O(\alpha_s)$ \cite{King:1986wi}. We will try to improve the situation by calculating all contributions at that order and presenting the results below.

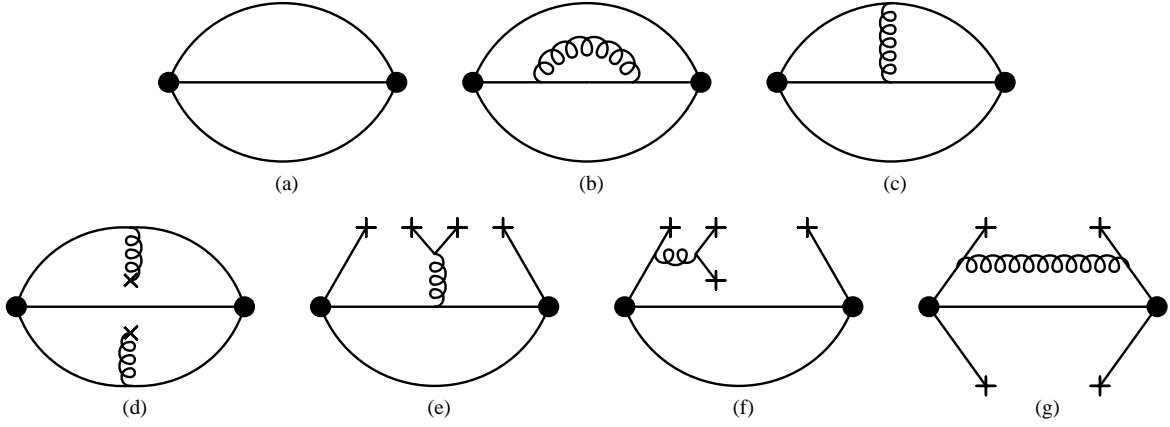
\begin{figure*}%
  \centering
  \begin{fmffile}{diagram}
  \unitlength 1mm
  \fmfset{curly_len}{2mm}
  \fmfstraight
  \subfloat[]{
    \begin{fmfgraph}(30,21)
        \fmfleft{i}
        \fmfright{o}
        \fmfv{d.s=c,d.si=7}{i,o}
          \fmf{plain,l=0.7}{i,o}
          \fmf{plain}{i,o}
          \fmf{plain,r=0.7}{i,o}
    \end{fmfgraph}
  }
  \qquad
  \subfloat[]{
    \begin{fmfgraph}(30,21)
        \fmfleft{i}
        \fmfright{o}
        \fmfv{d.s=c,d.si=7}{i,o}
          \fmf{plain,l=0.7}{i,o}
          \fmf{plain}{i,v1,vd,v2,o}
          \fmf{plain,r=0.7}{i,o}
          \fmf{gluon,r=0.7,t=0.25}{v2,v1}
    \end{fmfgraph}
  }
  \qquad
  \subfloat[]{
    \begin{fmfgraph}(30,21)
        \fmfleft{i}
        \fmfright{o}
        \fmftop{v1}
        \fmfv{d.s=c,d.si=7}{i,o}
          \fmf{plain,l=0.3}{i,v1,o}
          \fmf{plain}{i,v2,o}
          \fmf{plain,r=0.7}{i,o}
          \fmf{gluon,t=0}{v1,v2}
    \end{fmfgraph}
  }
  \\
  \subfloat[]{
    \begin{fmfgraph}(30,21)
        \fmfleft{i}
        \fmfright{o}
        \fmftopn{t}{11}
        \fmfbottomn{b}{11}
        \fmfv{d.s=c,d.si=7}{i,o}
        \fmfv{d.s=cr,d.si=7,d.a=0}{g1,g2}
          \fmf{plain,l=0.35}{i,t6,o}
          \fmf{plain}{i,o}
          \fmf{plain,r=0.35}{i,b6,o}
          \fmf{phantom}{t6,g1,g2,b6}
        \fmffreeze
          \fmf{gluon,t=0}{g1,t6}
          \fmf{gluon,t=0}{g2,b6}
    \end{fmfgraph}
  }
  \qquad
  \subfloat[]{
    \begin{fmfgraph}(30,21)
        \fmfleft{i}
        \fmfright{o}
        \fmftopn{t}{6}
        \fmfbottomn{b}{6}
        \fmfv{d.s=c,d.si=7}{i,o}
        \fmfv{d.s=cr,d.si=7,d.a=45}{t2,t3,t4,t5}
          \fmf{plain}{i,t2}
          \fmf{plain}{t5,o}
          \fmf{plain}{i,v1,o}
          \fmf{plain,r=0.7}{i,o}
        \fmffreeze
          \fmf{plain}{t3,v2,t4}
          \fmf{gluon}{v1,v2}
    \end{fmfgraph}
  }
  \qquad
  \subfloat[]{
    \begin{fmfgraph}(30,21)
        \fmfleft{i}
        \fmfright{o}
        \fmftopn{t}{6}
        \fmfbottomn{b}{6}
        \fmfv{d.s=c,d.si=7}{i,o}
        \fmfv{d.s=cr,d.si=7,d.a=45}{t2,t3,t5,p4}
          \fmf{plain}{i,vd,v1,t2}
          \fmf{plain}{t5,o}
          \fmf{plain}{i,o}
          \fmf{plain,r=0.7}{i,o}
          \fmf{phantom}{b3,p1,p2,p3,p4,p5,t3}
        \fmffreeze
          \fmf{plain}{t3,v2,p4}
          \fmf{gluon}{v1,v2}
    \end{fmfgraph}
  }
  \qquad
  \subfloat[]{
    \begin{fmfgraph}(30,21)
        \fmfleft{i}
        \fmfright{o}
        \fmftopn{t}{5}
        \fmfbottomn{b}{5}
        \fmfv{d.s=c,d.si=7}{i,o}
        \fmfv{d.s=cr,d.si=7,d.a=45}{t2,t4,b2,b4}
          \fmf{plain}{i,v1,t2}
          \fmf{plain}{t4,v2,o}
          \fmf{plain}{i,o}
          \fmf{plain}{i,b2}
          \fmf{plain}{b4,o}
          \fmf{gluon,t=0}{v2,v1}
    \end{fmfgraph}
  }
  \caption{\label{fig:diagrams}Feynman diagrams contributing to the CZ two-point function at $O(\alpha_s)$.}
  \end{fmffile}
\end{figure*}

We performed all calculations assuming massless up- and down-quarks and using dimensional regularization in the $\overline{\mathrm{MS}}$-scheme. Diagrammatical representations for all contributions to the two-point function \eqref{eq:2point} of two Chernyak-Zhitnitsky currents \eqref{eq:CZ} are given in Fig.~\ref{fig:diagrams}.
\section{\label{sec:sumrules}Sum rules}%
In order to extract values for the normalization constants we consider Borel-transformed sum rules \cite{Shifman:1978bx}. Sum rules are a standard technique whose details will not be explained here, see \cite{Colangelo:2000dp} for an introduction to the method. In Tab.~\ref{tab:results} we show detailed results where the common factor $(qz)\s{z}$ has been removed and the Borel-transformation
\begin{align}%
  \B \Bigl[f(q^2)\Bigr] = \lim_{\begin{subarray}{l}-q^2,n\to\infty \\ -q^2/n=M^2\end{subarray}} \frac{(-q^2)^{n+1}}{n!} \Biggl(\frac{d}{dq^2}\Biggr)^n f(q^2)
\end{align}%
has been performed. Each entry corresponds to a diagram shown in Fig.~\ref{fig:diagrams}.
\newcolumntype{L}{>{$\displaystyle}l<{$}}
\begin{table}%
  \caption{\label{tab:results}Contributions to the sum rule for CZ-currents.}
  \centering
  \begin{tabular}{cL}%
    \toprule
    (a) & \frac{1}{(2\pi)^4} \frac{1}{30} \int\limits_0^\infty ds \, s \, e^{-\frac{s}{M^2}}
    \\\addlinespace
    (b) & \frac{\alpha_s}{\pi} \frac{1}{(2\pi)^4} \int\limits_0^\infty ds \, s \Biggl(-\frac{197}{1800} -\frac{1}{30}\ln\frac{s}{\mu^2}\Biggr) e^{-\frac{s}{M^2}}
    \\\addlinespace
    (c) & \frac{\alpha_s}{\pi} \frac{1}{(2\pi)^4} \int\limits_0^\infty ds \, s \Biggl(\frac{407}{2700} +\frac{1}{45}\ln\frac{s}{\mu^2}\Biggr) e^{-\frac{s}{M^2}}
    \\\addlinespace
    (d) & \frac{\alpha_s}{\pi} \frac{1}{864\pi^2} \braket{G^2}
    \\\addlinespace
    (e) & \frac{\alpha_s}{\pi} \Biggl(\frac{2}{3^4}-\frac{2^2}{3^4}\Biggr) \frac{1}{M^2} \braket{\bar{q}q}^2
    \\\addlinespace
    (f) & \frac{\alpha_s}{\pi} \frac{2}{3^4} \frac{1}{M^2} \braket{\bar{q}q}^2
    \\\addlinespace
    (g) & \frac{\alpha_s}{\pi} \frac{2}{3^3} \frac{1}{M^2} \braket{\bar{q}q}^2
    \\\bottomrule
  \end{tabular}%
\end{table}%

All four-quark condensates have been reduced to the square of the two-quark condensate using the factorization hypothesis \cite{Shifman:1978bx} (with superscript color- and subscript Dirac-indices):
\begin{align}%
\begin{split}%
  \braket{\bar{q}_i^aq_j^b\bar{q}_k^cq_l^d} 
  &\simeq \braket{\bar{q}_i^aq_j^b}\braket{\bar{q}_k^cq_l^d} - \braket{\bar{q}_i^aq_l^d}\braket{\bar{q}_k^cq_j^b}
\\
  &= \frac{\braket{\bar{q}q}^2}{(4N_C)^2} \Bigl(\mathds{1}_{ij}\mathds{1}_{kl}\delta^{ab}\delta^{cd} - \mathds{1}_{il}\mathds{1}_{kj}\delta^{ad}\delta^{cb}\Bigr) ~.
\end{split}%
\end{align}%
The calculated condensate contributions have been found to be in agreement with \cite{King:1986wi}.

Using the results from Tab.~\ref{tab:results} and the hypothesis of quark-hadron duality, which introduces the effective threshold $s_0$, we postulate the sum rule for $f_N$ in next-to-leading order including operators up to dimension 6. The full sum rule reads:
\begin{align}%
\begin{split}%
\label{eq:nloresult}
  2\abs{f_N}^2 \; e^{-\frac{m_N^2}{M^2}} &= \frac{1}{(2\pi)^4} \; \frac{1}{30} \; \int\limits_0^{s_0} ds\; s\; e^{-\frac{s}{M^2}}
  \Biggl[1 + \frac{\alpha_s}{\pi}\Biggl(\frac{223}{180} - \frac{1}{3}\ln\frac{s}{\mu^2}\Biggr)\Biggr]
\\&
  \quad + \frac{\alpha_s}{\pi} \; \frac{1}{864\pi^2} \; \braket{G^2}
  + \frac{\alpha_s}{\pi} \; \frac{2}{27} \; \frac{1}{M^2} \; \braket{\bar{q}q}^2 ~,
\end{split}%
\intertext{%
which constitutes the main result of this work. In addition we also present the sum rule obtained from \cite{Sadovnikova:2005ye}:}
\begin{split}%
\label{eq:nlolambda}
  2 (2\pi)^4 \abs{\lambda_1}^2 \; m_N^2 \; e^{-\frac{m_N^2}{M^2}} &=
  M^6 E_3 \Biggl[1+\frac{\alpha_s}{\pi}\Biggl(\frac{53}{12}-\ln\frac{s_0}{\mu^2}\Biggr)\Biggr]
\\&\quad
  -\frac{\alpha_s}{\pi} \Biggl(M^4s_0\Biggl(1+\frac{3s_0}{4M^2}\Biggr)e^{-\frac{s_0}{M^2}}
  +M^6 \mathcal{E}\biggl(-\frac{s_0}{M^2}\biggr)\Biggr)
\\&
  \quad + \frac{b}{4} M^2 E_1
  + \frac{4}{3} a^2 \Biggl[
    1 - \frac{\alpha_s}{\pi}\Biggl(\frac{5}{6}+\frac{1}{3}\Biggl(\ln\frac{s_0}{\mu^2}+\mathcal{E}\biggl(-\frac{s_0}{M^2}\biggr)\Biggr)\Biggr)
    -\frac{1}{3}\frac{m_0^2}{M^2} \Biggr] ~.
\end{split}%
\intertext{%
The standard abbreviations used here are $a=-(2\pi)^2\braket{\bar{q}q}$, $b=(2\pi)^2\Braket{\frac{\alpha_s}{\pi}G^2}$, $m_0^2=\frac{\Braket{\bar{q}g\sigma Gq}}{\braket{\bar{q}q}}\approx\unit{(0.65-0.8)}{\giga\electronvolt\squared}$, $E_n=1-e^{-\frac{s_0}{M^2}}\;\sum_{i=0}^{n-1}\frac{1}{i!}\Bigl(\frac{s_0}{M^2}\Bigr)^i$. The function $\mathcal{E}$ is defined as $\mathcal{E}(x)=\sum_{n=1}^\infty\frac{x^n}{n\cdot n!}$. For better comparison with \eqref{eq:nloresult} we can rewrite the first part of \eqref{eq:nlolambda} in an integral form:%
}
\begin{split}%
  2 (2\pi)^4 \abs{\lambda_1}^2 \; m_N^2 \; e^{-\frac{m_N^2}{M^2}} &=
  \frac{1}{2} \; \int\limits_0^{s_0} ds\; s^2\; e^{-\frac{s}{M^2}}
  \Biggl[1 + \frac{\alpha_s}{\pi}\Biggl(\frac{71}{12} - \ln\frac{s}{\mu^2}\Biggr)\Biggr]
\\&
  \quad + \frac{b}{4} M^2 E_1
  + \frac{4}{3} a^2 \Biggl[
    1 - \frac{\alpha_s}{\pi}\Biggl(\frac{5}{6}+\frac{1}{3}\Biggl(\ln\frac{s_0}{\mu^2}+\mathcal{E}\biggl(-\frac{s_0}{M^2}\biggr)\Biggr)\Biggr)
    -\frac{1}{3}\frac{m_0^2}{M^2} \Biggr] ~.
\end{split}%
\end{align}%
\section{\label{sec:analysis}Numerical analysis}%
We can now proceed to analyze the new sum rule \eqref{eq:nloresult}. For comparison we will also consider the next-to-leading order sum rule for the Ioffe current which has been derived in \cite{Sadovnikova:2005ye}.

\begin{figure*}[p]%
  \centering
  \subfloat[Analysis of the sum rule for $\abs{f_N}$]
    {\includegraphics[height=0.25\textheight]{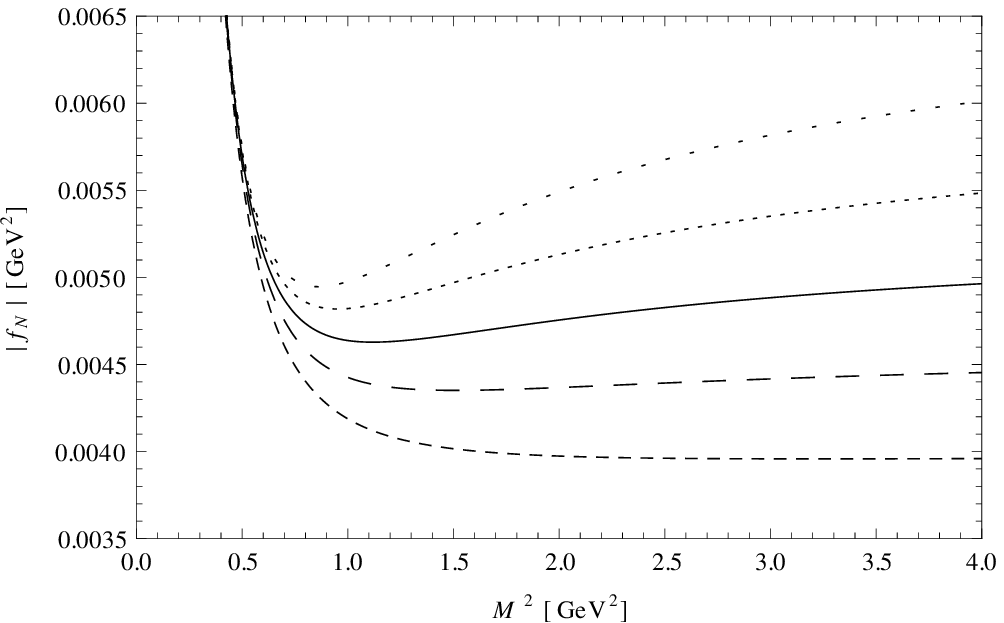}}\\
  \subfloat[Analysis of the sum rule for $\abs{\lambda_1}$]
    {\includegraphics[height=0.25\textheight]{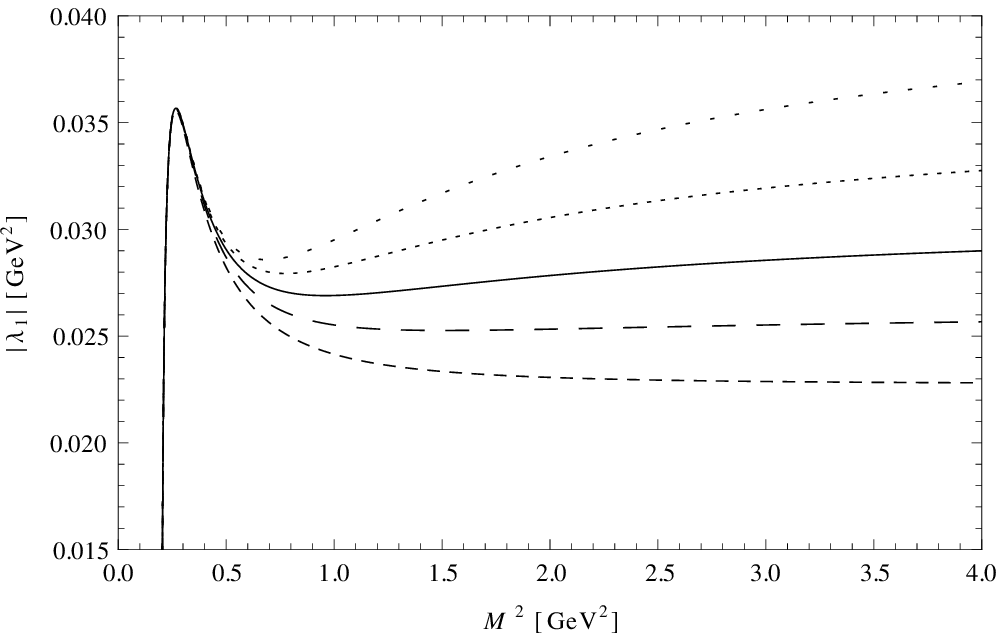}}\\
  \subfloat[Analysis of the ratio $\abs{f_N}/\abs{\lambda_1}$]
    {\includegraphics[height=0.25\textheight]{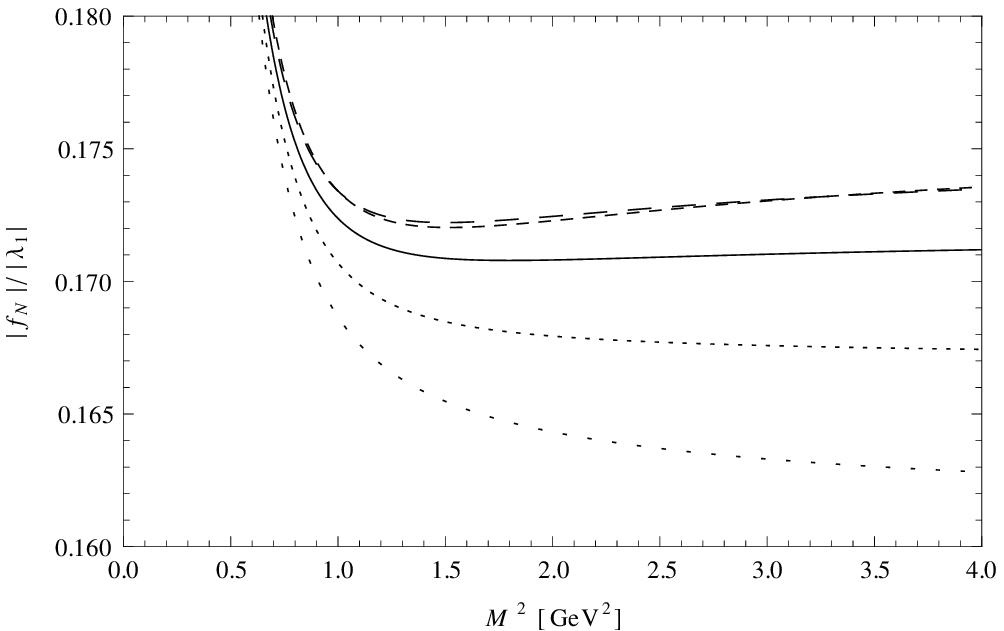}}
  \caption{\label{fig:loplots}Plots for the coupling constants $f_N$ and $\lambda_1$ in leading order at $s_0=(\unit{1.3}{\giga\electronvolt})^2$(short dashed), $(\unit{1.4}{\giga\electronvolt})^2$(long dashed), $(\unit{1.5}{\giga\electronvolt})^2$(solid), $(\unit{1.6}{\giga\electronvolt})^2$(dotted) and $(\unit{1.7}{\giga\electronvolt})^2$(light dotted).}
\end{figure*}

\begin{figure*}[p]%
  \centering
  \subfloat[Analysis of the sum rule \eqref{eq:nloresult} for $\abs{f_N}$]
    {\includegraphics[height=0.25\textheight]{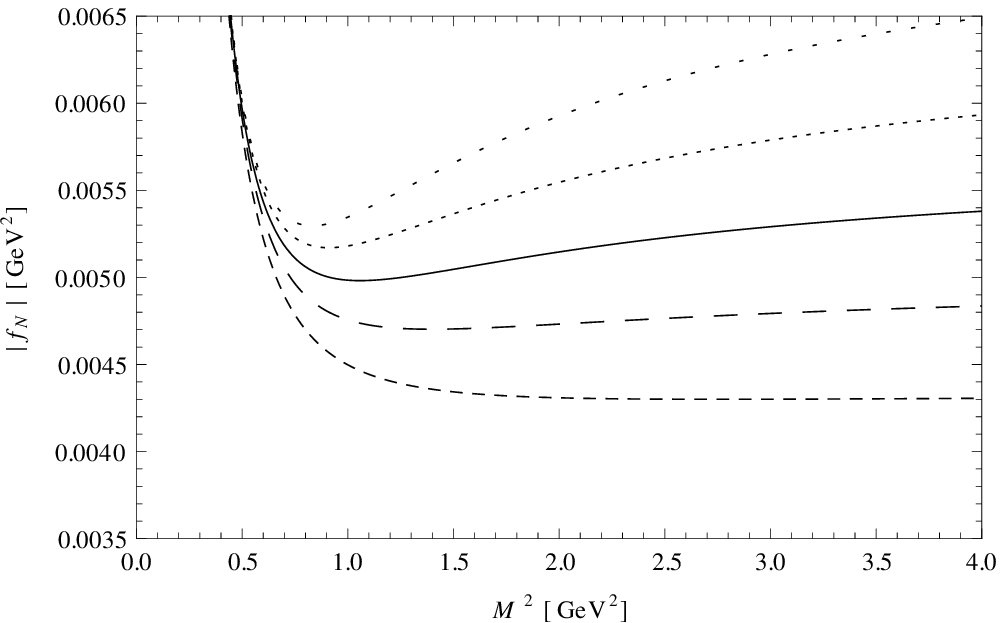}}\\
  \subfloat[Analysis of the sum rule \eqref{eq:nlolambda} for $\abs{\lambda_1}$]
    {\includegraphics[height=0.25\textheight]{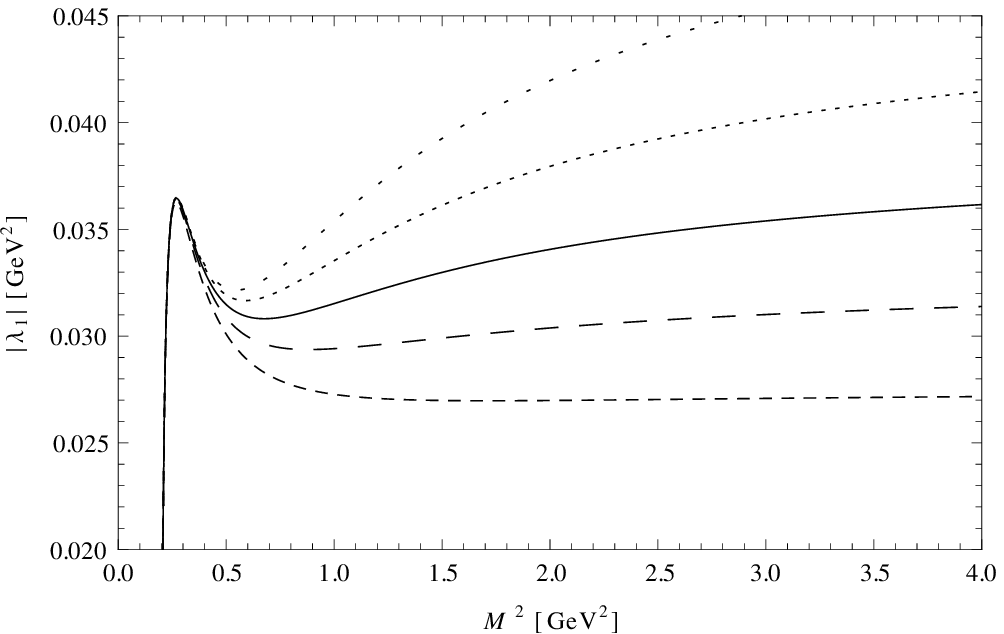}}\\
  \subfloat[Analysis of the ratio $\abs{f_N}/\abs{\lambda_1}$]
    {\includegraphics[height=0.25\textheight]{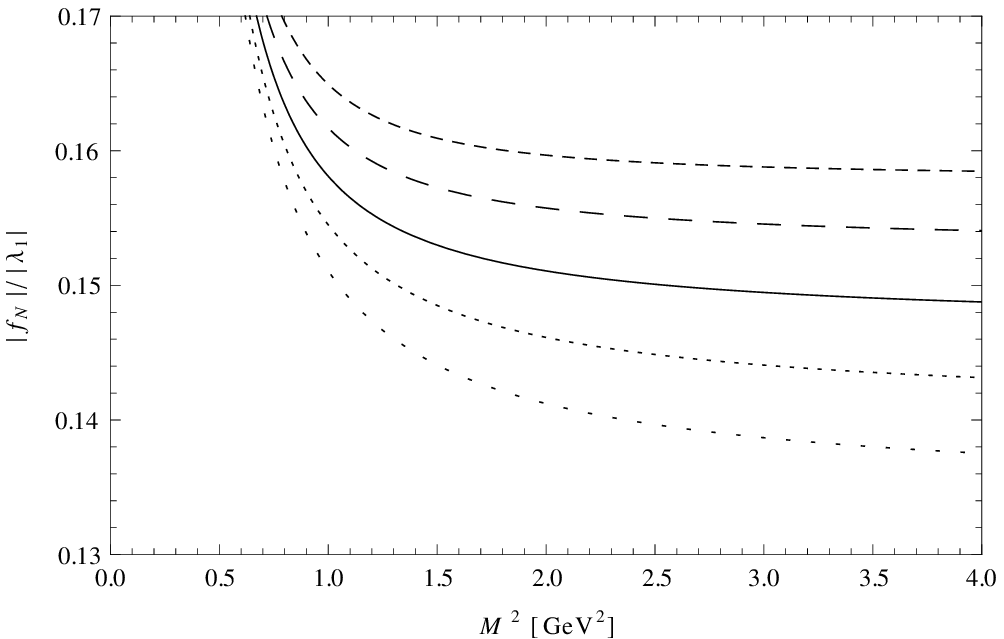}}
  \caption{\label{fig:nloplots}Plots for the coupling constants in next-to-leading order at $s_0=(\unit{1.3}{\giga\electronvolt})^2$(short dashed), $(\unit{1.4}{\giga\electronvolt})^2$(long dashed), $(\unit{1.5}{\giga\electronvolt})^2$(solid), $(\unit{1.6}{\giga\electronvolt})^2$(dotted) and $(\unit{1.7}{\giga\electronvolt})^2$(light dotted).}
\end{figure*}

Let us consider the normalization constants obtained from the sum rules both at leading order and at next-to-leading order, shown in Fig.~\ref{fig:loplots} and Fig.~\ref{fig:nloplots}, respectively, as a function of the Borel parameter $M^2$. The numerical values of the other parameters have been chosen as follows: The renormalization scale has been fixed at $\mu_{\overline{\mathrm{MS}}}=\unit{1}{\giga\electronvolt}\simeq m_N$. For the value of the quark condensate we used a recent determination ($\mu=\unit{1}{\giga\electronvolt}$) \cite{Jamin:2002ev}
\begin{align}%
\label{eq:quarkcondensate}
  \braket{\bar{q}q} &= -(\unit{242 $\pm$ 15}{\mega\electronvolt})^3 ~,
\intertext{while a standard value has been used for the gluon condensate:}
\label{eq:gluoncondensate}
  \Braket{\frac{\alpha_s}{\pi}G^2} &= \unit{(0.012 $\pm$ 0.006)}{\power{\giga\electronvolt}{4}} ~.
\end{align}%
The threshold parameter has been varied around $\sqrt{s_0}\simeq\unit{1.5}{\giga\electronvolt}$, which is a standard choice due to being of the same size as the mass of the lowest hadronic resonances. The leading order sum rules have been found to be quite stable in this region, though the next-to-leading order sum rules turn out to be a little less stable, cf. Fig.~\ref{fig:loplots} and Fig.~\ref{fig:nloplots}. As one can see from the plots the next-to-leading order correction is positive for both $\abs{f_N}$ and $\abs{\lambda_1}$.

Tab.~\ref{tab:constants} summarizes the various values for the normalization constants. For the sum rules the values have been extracted from the Borel window $\unit{1}{\giga\electronvolt\squared}\lesssim M^2\lesssim\unit{3}{\giga\electronvolt\squared}$. The errors are estimated by varying the nonperturbative parameters according to \eqref{eq:quarkcondensate} and \eqref{eq:gluoncondensate}, incorporating a $10\%$ uncertainty for $\alpha_s(\unit{1}{\giga\electronvolt})\approx0.5$ and also varying the threshold $\sqrt{s_0}$ by $10\%$ (i.e.\ adjusting $s_0$ by $20\%$). Other sources of error such as NNLO corrections, operators of higher dimension, condensate factorization, etc.\ are not easily estimated and thus not included. For the lattice simulation the quoted errors cover the combined statistical and systematic uncertainties given in\ \cite{Braun:2008ur}.

It can be observed that our NLO correction to $\abs{f_N}$ amounts to $\sim10\%$, while $\abs{\lambda_1}$ is increased by as much as $\sim20\%$. It should be noted that the new value for $\abs{f_N}$ of $\unit{5.1\cdot\power{10}{-3}}{\giga\electronvolt\squared}$ does not coincide with recent lattice simulations on that matter, which put the value at about $\unit{3.2\cdot\power{10}{-3}}{\giga\electronvolt\squared}$.
\newcolumntype{C}{>{$}c<{$}}
\begin{table}%
  \caption{\label{tab:constants}Values for the normalization constants obtained from the sum rules \eqref{eq:nloresult} and \eqref{eq:nlolambda} in leading and next-to-leading order compared to lattice results from \cite{Braun:2008ur} (at $\mu=\unit{1}{\giga\electronvolt}$).}
  \centering
  \begin{tabular}{rCCC}%
    \toprule
    &\abs{f_N}~[\giga\electronvolt\squared]&\abs{\lambda_1}~[\giga\electronvolt\squared]&\abs{f_N}/\abs{\lambda_1}
    \\\midrule
    LO   &  (4.7\pm0.7)\cdot10^{-3}  &  (2.8\pm0.6)\cdot10^{-2}  &  1.7\cdot10^{-1}
    \\
    NLO  &  (5.1\pm0.8)\cdot10^{-3}  &  (3.4\pm0.8)\cdot10^{-2}  &  1.5\cdot10^{-1}
    \\
    Lat  &  (3.2\pm0.2)\cdot10^{-3}  &  (3.6\pm0.2)\cdot10^{-2}  &  0.9\cdot10^{-1}
    \\\bottomrule
  \end{tabular}%
\end{table}%

The discussion above only covered the magnitudes of the normalization constants. Their phases cannot be fixed, since the sum rules only depend on the square of the absolute value. However, we can make a statement about the relative phase between $f_N$ and $\lambda_1$. By considering a non-diagonal correlator (between $\eta_{CZ}$ and $\eta_I$) and taking a ratio of sum rules it can be determined that the ratio $f_N/\lambda_1$ is a negative real number \cite{Braun:2000kw}. A standard choice is to designate $f_N$ as positive and $\lambda_1$ as negative.
\section{\label{sec:conclusion}Conclusion}%
The next-to-leading order perturbative correction to the two-point function of leading-twist currents has been calculated in this work. Based on this correlator an improved SVZ sum rule was proposed and used to obtain a new value for the leading-twist normalization constant $f_N$. The leading order result is in agreement, within errors, with previous determinations \cite{Chernyak:1984bm,King:1986wi,Braun:2006hz}, the next-to-leading order corrections increase the size of $f_N$ by approximately $10\%$.

The correction turned out to be less significant than in the case of the coupling constant $\lambda_1$ which is associated with the Ioffe current. For this current the next-to-leading order corrections had been calculated before and the established sum rules have been analyzed in comparison. There it has been observed that the $O(\alpha_s)$ contributions raise the magnitude of $\lambda_1$ by approximately $20\%$.

The discrepancies between the sum rule estimations and lattice simulations could not be resolved. The calculated correction causes an increase of $f_N$, whereas the results obtained on the lattice \cite{Braun:2008ur} are lower than the traditional sum rule predictions. However, the lattice results have not yet stabilized and suffer from uncertainties. These originate, for instance, from chiral extrapolation, which is due to the fact that the pion-mass dependence of $f_N$ is not fully understood.
\section*{Acknowledgements}%
The author wishes to thank V.~M.~Braun for his guidance and expertise and M.~Jamin for his hospitality and helpful discussions. This work was supported in part by the DAAD PPP grant D/07/13355.
%
%
%
\bibliographystyle{h-physrev5}
\bibliography{NSR}
\end{document}